\documentstyle{article}
\title{
\begin{flushright}
{\bf\normalsize   COLO-HEP-276}\\
\end{flushright}
\bf Multiple Potts Models Coupled to Two-Dimensional Quantum Gravity.
}

\author{ {\it C.F. Baillie} \\
         Physics Dept. \\
         University of Colorado\\
         Boulder, CO 80309, USA\\
	 \\
         and \\
	 \\
         {\it D.A. Johnston}\\
         Dept. of Mathematics\\
         Heriot-Watt University\\
         Riccarton\\
         Edinburgh, EH14 4AS, Scotland}

\begin{document}
  \maketitle
                      {\Large
                      \begin{abstract}
%
We perform Monte Carlo simulations using the Wolff cluster algorithm
of {\it multiple} $q=2,3,4$ state Potts models on dynamical phi-cubed graphs of
spherical topology
in order to investigate the $c>1$ region of two-dimensional quantum gravity.
Contrary to naive expectation we find no obvious signs of pathological
behaviour for $c>1$.
We discuss the results in the light of suggestions that have been made for a
modified
DDK ansatz for $c>1$.
\\
\\
Submitted to Phys. Lett. B
%
                        \end{abstract} }
%
  \thispagestyle{empty}
%
%
  \newpage
%
                  \pagenumbering{arabic}

\section{Introduction}

In a recent paper \cite{1}, we carried out numerical simulations
of the $q=2$(Ising) and $q=3,4,10$ state Potts models on
dynamical phi-cubed graphs of spherical topology in order
to see if the predictions of KPZ \cite{2} and DDK \cite{3}
for the critical exponents of the $q=2,3,4$ models were in agreement with
``experiment''.
We found that the predicted and measured exponents were in quite good agreement
for these models and that the $q=10$ Potts model still displayed a first order
transition on
the dynamical phi-cubed graph as it did on a fixed lattice. We also explored
the intrinsic geometry of
the graphs in the simulation, finding a linear relation between the number of
rings of length three at the critical temperature and the central charge, and
that the peak in the ring distribution moved towards $\beta_c$ as $q$ increased
\footnote{In \cite{1} we stated that the peak moved towards $\beta_c$ as $c$
increased,
as $c$ increases with $q$ for the single Potts models.
We shall see in this paper that this is not strictly accurate
for multiple Potts models.}.
In the current paper we employ the same methods to explore {\it multiple}
copies
of $q=2,3,4$ state Potts models on dynamical phi-cubed graphs.
We shall simulate $2,4,8$ and $16$ Ising models, $5$ and $10$ $q=3$ Potts
models, and
$2,4$ and $8$ $q=4$ Potts models all of which have, at least naively, $c \ge
1$.

The formulae of KPZ/DDK give nonsensical results for $c >1$, suggesting there
is some sort of phase transition at $c=1$. The results of simulations
of dynamically triangulated random surfaces, where the phi-cubed graphs
or their dual triangulations are embedded in $d \ge 3$ ($c=d$ for the
Polyakov action for a surface) dimensions seem to support
this view since they give rise to very crumpled and spiky surfaces with
no continuum limit \cite{4}.
We can include an extrinsic curvature squared \cite{5}, or mean curvature
\cite{5a},
term in addition to the Polyakov action
which may cure these ills.
There have been suggestions \cite{6} that there is a
liberation of curvature singularities at $c=1$ when such a term is not included
which give rise to the spikiness
observed in simulations. As this is essentially an intrinsic phenomenon we
would
expect to see some sort of pathology for {\it any} model on a dynamical
triangulation
(or its dual) whose naive continuum limit, if it existed at all, had $c >1$.
The
simplest sort of matter we can put on a dynamical graph
is a spin variable, as in the various Potts models, and we
can arrange $c \ge 1$ by taking multiple copies.
The copies interact with each other only through the graph.
The simplicity of the matter
allows us to explore quite large graphs by comparison with simulations of
embedded
surfaces. As the graphs are no longer embedded the pathology, if any, cannot
manifest itself as ``crumpling''. The object of the simulations described
in this paper is to see if things go wrong in some other manner for models with
$c$ (naively) $ >1$.

The partition function we are simulating is
that for multiple Potts models on dynamical random graphs with a fixed number
of nodes
$N$ \cite{6a} and is given by
\begin{equation}
Z_N =  \sum_{G^{(N)}} \sum_{\sigma} \exp \left( - {\beta \over 2} \sum_{\alpha}
\sum_{i,j=1}^N G^{(N)}_{ij}
\delta( \sigma_i^{\alpha}  \sigma_j^{\alpha}) \right)
\label{e2}
\end{equation}
where
$G^{(N)}$ is the adjacency matrix of the phi-cubed graph,
$\delta$ is a Kronecker delta,
$\beta$ is the inverse temperature $1/T$ and the sum over $\alpha$ is a sum
over the
different copies of the Potts model.
We have, in general, $q$ species of spin taking the values $0,1,...,q-1$.
A single Ising model has been simulated on both dynamical triangulations
\cite{7},\cite{8} and phi-cubed graphs \cite{9}
and satisfactory agreement between the
measured and theoretical values of the critical exponents found.
The theoretical values come from both the work in \cite{2},\cite{3}, as
outlined below, and
the exact solution \cite{10}.
If we denote the critical temperature for a continuous spin-ordering
phase transition by $T_c$ and the reduced temperature $|T - T_c|/T_c$ by $t$
then the
critical exponents $\alpha, \beta, \gamma, \nu, \delta, \eta$
can be defined in the standard manner as $t \rightarrow 0$
\begin{eqnarray}
C \simeq t^{-\alpha} \; &;& \; M \simeq t^{\beta}, T<T_c \nonumber \\
\chi \simeq t^{- \gamma} \; &;& \;
\xi \simeq t^{- \nu} \nonumber \\
M(H,t=0) &\simeq& H^{1 / \delta}, \; H \rightarrow 0 \nonumber \\
<M(x) M(y)> &\simeq& {1 \over |x - y |^{d -2 + \eta}}, \; t=0
\label{e04}
\end{eqnarray}
where $C$ is the specific heat, $M$ is the magnetization,
$\chi$ is the susceptibility, $\xi$ is the correlation length and
$H$ is an external field.
In theories without gravity
it is possible to calculate $\alpha$ and $\beta$
using the conformal weights of the energy density operator and spin operator
(for a review see \cite{11a}).
Given these
we can now use the various scaling relations \cite{11b}
\begin{eqnarray}
\alpha &=& 2 - \nu d \nonumber \\
\beta  &=& {\nu \over 2} (d - 2 + \eta) \nonumber \\
\gamma &=& \nu ( 2 - \eta) \nonumber \\
\delta &=& { d + 2 - \eta \over d - 2 + \eta }
\label{e033}
\end{eqnarray}
to obtain the other exponents.

Miraculously, this approach carries over directly
when one considers the same theories coupled to gravity (i.e. on dynamical
meshes):
the modified conformal weights $\Delta$ for the energy and spin operators  when
coupled to
gravity are given in \cite{2},\cite{3} in terms of the original
conformal weights $\Delta_0$ by
\begin{equation}
\Delta - \Delta_0 = - {\xi^2 \over 2} \Delta ( \Delta - 1),
\label{e0a}
\end{equation}
where
\begin{equation}
\xi = - { 1 \over 2 \sqrt{3} } ( \sqrt{ 25 -c } - \sqrt{ 1 -c} ),
\label{e0b}
\end{equation}
and these simply replace the standard
values in the calculation of the critical exponents $\alpha$ and $\beta$. The
one exact solution
that is currently available, for the Ising model on a dynamical mesh, has
exponents
which satisfy the scaling relations so we assume their validity to derive the
other exponents from these
two.

In \cite{1} we found that there was satisfactory agreement with the results of
\cite{2},\cite{3}
for single copies of the $q=3,4$ state Potts models and that the $q=10$ Potts
models had a
first order transition. For single Potts
models this suggests the following appealing picture:
for $q \le 4$ they display a continuous phase transition, which allows
us to define a continuum limit, whereas the $q > 4$ models, as on  fixed
lattices, have
first order transitions with no continuum limit and associated conformal field
theory.
It is also tempting to speculate that the difficulties that appear in the
continuum formalism
for $c>1$ may be due to a similar effect, with multiple copies of the Potts
models having
first order transitions when $c>1$. In what follows we shall see that the
simulations
fail to oblige our expectations.

\section{Spin Model Properties}

We perform a microcanonical (fixed number of nodes)
Monte Carlo simulation on phi-cubed graphs with
$N = 50,100,200,300,500,1000$ and $2000$ nodes
at various values of $\beta$ between $0.1$ and $1.5$.
The Monte Carlo update consists of two parts:
one for the spin model and one for the graph.
For the Potts model we use a cluster update algorithm
since it suffers from much less critical slowing down than the
standard Metropolis algorithm.
There are two popular cluster algorithm
implementations - Wolff \cite{12} and Swendsen-Wang \cite{13}.
As they are equivalent for the usual two-dimensional $q=2,3$ Potts
models we use Wolff's variant because it is computationally faster.
For the graph update we use the Metropolis algorithm with the
standard ``flip'' move.  As we are working with phi-cubed
graphs the detailed balance condition involves checking that the rings
at either ends of the link being flipped have no links in common.
This check also eliminates all graphs containing tadpoles or self-energies.
After each Potts model update sweep we randomly pick $NFLIP$ links one
after another and try to flip them. After testing various values of $NFLIP$
to ensure that there were enough flips to make the graph dynamical on the
time scale of the Potts model updates
we set $NFLIP=N$ for all the simulations.

Just as in our simulations of
single Potts models we measure all the standard thermodynamic quantities for
the spin model:
energy $E$, specific heat $C$, magnetization $M$, susceptibility $\chi$
and correlation length $\xi$; and several properties of the graph:
acceptance rates for flips, distribution of ring lengths and
internal fractal dimension $d$.
To determine $\nu$ (actually $\nu d$) and $\beta_c$
separately, instead of from the usual three-parameter
finite-size scaling fit, for example $\xi = \xi_0 (|T-T_c|/T_c)^{-\nu}$,
we used Binder's cumulant \cite{14}.
This is done as follows.
Binder's cumulant $U_N$ on graph with $N$ nodes is defined as
\begin{equation}
U_N = 1 - {<M^4> \over 3<M^2>^2},
\end{equation}
where $<M^4>$ is the average of the fourth power of the magnetization
and $<M^2>$ is the average of its square.
For a normal temperature-driven continuous phase transition
$U_N \rightarrow 0$ for $T > T_c$ because $M$ is
gaussian distributed about 0 at high temperature,
and $U_N \rightarrow {2 \over 3}$
for $T < T_c$  because a spontaneous magnetization $M_{sp}$
develops in the low temperature phase.
At $T = T_c$, $U_N$ has a non-trivial value which scales with $N$ according to
\begin{equation}
U_N \simeq t N^{1 \over \nu d}.
\end{equation}
Therefore the slope of $U_N$ with respect to $T$ (or $\beta$) at $T_c$
gives ${1 \over \nu d}$.
This is not much use as it stands since it involves knowledge of $T_c$,
but the $maximum$ value of the slope
scales in the same way, so we can extract $\nu d$ from
\begin{equation}
\max({dU_N \over d\beta}) \simeq N^{1 \over \nu d}.
\end{equation}
We show the results of doing this in Figs. 1a,b,c for the $q=2,3,4$ state Potts
models
respectively. From the figures it is clear that there is little variation
in the values of $\nu d$ for a given $q$ as the number of copies of the spin
model is increased. We therefore use a {\it weighted average} $\nu d$ for each
$q$
in what follows: 3.6(4) for $q=2$, 2.9(4) for $q=3$, and 2.8(2) for $q=4$.

We now make use of these values of $\nu d$ in the standard finite-size scaling
relation
(with $L$ replaced by $N^{1 \over d}$ since we do not know $d$ a priori)
\begin{equation}
|\beta_c^N - \beta_c^\infty| \simeq N^{-{1 \over \nu d}}
\end{equation}
to extract $\beta_c^\infty$, using $\beta_c^N$s obtained from
the position of the maximum in the slope of $U_N$ or from the
peak in the specific heat $C$. The end results are shown in Table 1, where
we list the model in the form `number of copies' q-value, central charge $c$,
$\nu d$,
$\beta_c$ estimated from $U_N$ and $\beta_c$ estimated from $C$. It is clear
that
the peak in the specific heat gives the more accurate estimates for $\beta_c$,
and
we use this in the finite size scaling analysis which follows.

\def\bc{\beta_c^\infty}

\begin{center}
\begin{tabular}{|c|c|c|c|c|} \hline
model & $c$ & $\nu d$ & $\bc (U_N)$ & $\bc (C)$ \\[.05in]
\hline
 2 q2 &  1  & 3.5(3)  & 0.80(5)     & 0.79(1)   \\[.05in]
 4 q2 &  2  & 3.9(4)  & 0.87(4)     & 0.82(1)   \\[.05in]
 8 q2 &  4  & 3.8(4)  & 0.84(4)     & 0.84(1)   \\[.05in]
16 q2 &  8  & 3.4(5)  & 0.92(5)     & 0.88(1)   \\[.05in]
\hline
 5 q3 &  4  & 2.9(2)  & 0.93(3)     & 0.91(1)   \\[.05in]
10 q3 &  8  & 3.2(5)  & 0.99(3)     & 0.96(1)   \\[.05in]
\hline
 2 q4 &  2  & 2.8(2)  & 0.96(3)     & 0.95(1)   \\[.05in]
 4 q4 &  4  & 2.9(2)  & 0.97(3)     & 0.96(1)   \\[.05in]
 8 q4 &  8  & 2.8(1)  & 1.02(3)     & 0.99(1)   \\[.05in]
\hline
\end{tabular}
\end{center}
\vspace{.1in}
\centerline{Table 1: Fitted values of $\nu d$ and inverse critical temperature
$\beta_c$}
\centerline{from Binder's cumulant $U_N$ and specific heat $C$.}
\bigskip

We have chosen this slightly non-standard approach to obtaining $\beta_c$,
rather than looking
for the crossing of cumulants on different sized meshes, because we found it
very difficult
to judge the crossing points accurately. Applying our approach to the single
Ising model
in \cite{1} gave an accurate estimate of its known $\beta_c$. A comparison of
the results of multiple
Potts models with those for single models in \cite{1} shows that $\beta_c$
increases
slowly with the number of models, but that $\nu d$ stays roughly constant at a
value
somewhat larger than that for a single model.

We can now measure the other critical exponents
either from the singular behaviour of the thermodynamic functions
\begin{equation}
C = B + C_0 t^{-\alpha}, \; M = M_0 t^{\beta}, \; \chi = \chi_0 t^{-\gamma},
\; \xi = \xi_0 t^{-\nu}
\label{eF1}
\end{equation}
knowing $\beta_c$,
or from the finite-size scaling relations
\begin{equation}
C = B' + C_0' N^{\alpha \over \nu d}, \; M = M_0' N^{-\beta \over \nu d}, \;
\chi = \chi_0' N^{\gamma \over \nu d}
\label{eF2}
\end{equation}
using the previously obtained values of $\nu d$. The results obtained for
$\beta / \nu d$, $\gamma / \nu d$, $\gamma$ and $\nu$
in the various models obtained from the above fits
are shown in Table 2

\begin{center}
\begin{tabular}{|c|c|c|c|c|c|c|c|} \hline
model &$c$&$\beta/\nu d$& $\gamma/\nu d$ & $\gamma$ & $\nu$   & $d$ \\[.05in]
\hline
 2 q2 & 1 & 0.11(1)     & 0.83(1)        & 2.4(1)  & 0.87(2) & 2.33(4)
\\[.05in]
 4 q2 & 2 & 0.08(1)     & 0.94(2)        & 3.1(2)  & 0.93(4) & 2.75(4)
\\[.05in]
 8 q2 & 4 & 0.09(1)     & 0.98(1)        & 3.3(2)  & 0.83(2) & 2.67(4)
\\[.05in]
16 q2 & 8 & 0.08(1)     & 1.05(2)        & 2.7(1)  & 1.02(2) & 2.59(4)
\\[.05in]
\hline
 5 q3 & 4 & 0.09(1)     & 0.91(1)        & 2.2(1)  & 0.97(2) & 2.58(3)
\\[.05in]
10 q3 & 8 & 0.05(1)     & 1.01(2)        & 2.8(2)  & 1.24(2) & 2.56(3)
\\[.05in]
\hline
 2 q4 & 2 & 0.09(1)     & 0.86(1)        & 2.0(1)  & 1.16(3) & 2.63(3)
\\[.05in]
 4 q4 & 4 & 0.11(1)     & 0.90(1)        & 1.7(1)  & 0.85(2) & 2.52(3)
\\[.05in]
 8 q4 & 8 & 0.16(1)     & 0.98(1)        & 2.9(1)  & 0.85(4) & 2.47(3)
\\[.05in]
\hline
\end{tabular}
\end{center}
\vspace{.1in}
\centerline{Table 2: Measured values of critical exponents $\beta, \gamma,
\nu$}
\centerline{and internal fractal dimension $d$ from $N=2000$ graphs.}

\bigskip

We see immediately that $\gamma / \nu d$ increases for $q=2,3,4$ with the
number of models,
whereas $\beta / \nu d$ decreases for $q=2,3$ and increases for $q=4$. However,
the measurements
of $\beta / \nu d$ proved to vary the most from the predicted values of KPZ for
the single
potts models in \cite{1}, so one should really conduct simulations on larger
meshes
before taking the trends in $\beta / \nu d$ at face value.
In order to make a meaningful comparison of these results with those for
single Potts models in \cite{1},
where $\gamma / \nu d$ {\it decreases} with $q$ and
$\beta / \nu d$ {\it increases} with $q$, we should look at multiple models
with
the {\it same} $c$.
Then we see that for $c=2$ $\gamma / \nu d$ decreases
from $0.94(2)$ for $q=2$ to $0.86(1)$ for $q=4$, and
$\beta / \nu d$ increases
from $0.08(1)$ for $q=2$ to $0.09(1)$ for $q=4$.
Similarly, for $c=4$ as $q$ increases from $2$ to $3$ to $4$
$\gamma / \nu d$ decreases from $0.98(1)$ to $0.91(1)$ to $0.90(1)$, and
 $\beta / \nu d$ increases from $0.09(1)$ to $0.09(1)$ to $0.11(1)$.
And finally for $c=8$
$\gamma / \nu d$ decreases from $1.05(2)$ to $1.01(2)$ to $0.98(1)$, and
 $\beta / \nu d$ increases from $0.08(1)$ to $0.05(1)$ to $0.16(1)$.
{}From these trends we would argue that the value of $\beta / \nu d$
for 10 q3 is a little low, but otherwise things look consistent.

The direct fits to $\gamma$ are poorer than those to $\beta / \nu d$ and
$\gamma / \nu d$ due to the uncertainty in $\beta_c$. In fact, if we increase
$\beta_c$ from the values given in the last column of Table 1 by the size
of the error bar ($0.1$) and re-do the fits we obtain values for $\gamma$
which are $0.2$ lower than those given in Table 2. Therefore our
values for $\gamma$ can be trusted only to about $10\%$.
These values are all larger than those obtained for single Potts models --
which makes sense as $\gamma$ controls the divergence of the
susceptibility and obviously with more spin models this divergence will be
stronger.
For single Potts models \cite{1} we know that $\gamma$ decreases with $q$
so as above we should compare our multiple models with the same $c$.
For $c=2$ and $c=4$ we find decreasing values but for $c=8$ we have
$\gamma=2.7,2.8,2.9$ for $q=2,3,4$ respectively. We could say that
this casts doubt on the fits for 10 q3 and 8 q4 but given the errors
judgement should be reserved until the next generation of simulations is
performed.
As a consistency check we can multiply our fit values of $\gamma / \nu d$
by $\nu d$ and see how close we get to the fitted values for $\gamma$.
Surprisingly, almost all results are within $10\%$, and the two
worst (16 q2 and 4 q4) only differ by $30\%$.

We have not attempted to fit the exponent $\beta$ directly here since
even for single Potts models we never managed to obtain satisfactory
results for this \cite{1}.

We fitted the exponent $\nu$ from the power law
divergence of the correlation length $\xi$ at $\beta_c$.
However, $\xi$ itself must be obtained from a fit:
the 2-point correlation function $\Gamma$ should behave as
\begin{equation}
\Gamma(r) \ \equiv \ \sum \sigma_i \sigma_{i+r} \ = \ c \ e^{- m r},
\label{eXI}
\end{equation}
where $m \equiv 1/\xi$, and the sum is over some number of measurements
made on each graph with the position of the spin $\sigma_i$ being chosen
randomly.
$r$ is the internal distance between two spins on the graph, i.e. the
fewest links between them.
As two fits are involved, both requiring knowledge of $\beta_c$,
the results are not particularly reliable.
If we re-do the fits with $\beta_c$ changed to $\beta_c+0.1$ then the sixth
column of Table 2 would read instead
$1.01(2), 1.11(5), 1.00(2), 1.19(2); 1.15(3), 1.55(2); 1.45(4), 1.04(2),
1.10(2)$.
Combining both sets of fits (we do not consider fits using
$\beta_c-0.01$ since our other estimates of $\beta_c$ from Binder's cumulant
are always {\it larger} than those from the specific heat)
yields our best estimates for these $\nu$ values as
$1.0(1), 1.0(1), 0.9(1), 1.1(1); 1.0(1), 1.4(1); 1.3(1), 1.0(1), 1.0(1)$.
Thus we conclude that $\nu$ is $1$ to an accuracy of $10\%$ for most
if not all the models.

As a further
consistency check one can also measure the internal fractal dimension $d$
directly,
as we discuss in the next section,
and see how the product of the (less accurate) individual measurements of $\nu$
and
$d$ compares with the direct determination of $\nu d$ from Binder's cumulant.
We can see by comparing Tables 1 and 2 that the product
underestimates $\nu d$ for almost all the models, but still appears to be
fairly good considering the multiple fits involved.

Unfortunately, due to the extra adjustable constants ($B,B'$)
in eqs. \ref{eF1},\ref{eF2} for the specific heat,
we are unable to make unambiguous fits for $\alpha$ and $\alpha / \nu d$.
Although we cannot fit directly for $\alpha$ we can use our
values of $\nu d$ and {\it assume} that the scaling relation
$\alpha = 2 - \nu d$ is still valid, which gives (with the
averaged value of $\nu d$ for a given $q$) $\alpha = -1.6(4)$ for $q=2$,
$\alpha = -0.9(4)$ for $q=3$ and $\alpha = -0.8(2)$ for $q=4$.
We thus see that this gives third order transitions for all $q$,
which are becoming weaker with increasing $q$.
These results should be compared with
the single Potts models where we have (analytically) $\alpha = -1$ for $q=2$,
$\alpha = -0.5$
for $q=3$ and $\alpha=0$ for $q=4$. Adding multiple copies
at a given $q$ thus appears to make the third order transitions stronger for
$q=2,3$
and change the borderline $q=4$ case from logarithmic second order to third
order.
In addition, visual inspection of the graphs of specific heat for various
$N$ suggests that they are not diverging with $N$, as one might expect
for a higher than second order transition.
We can, of course, rule
out a first order transition because we do not observe any discontinuities
in the energy or magnetization.

\section{Graph Properties}

We are particularly interested in the properties of the graphs when $c>1$ to
see if there
are any indications of pathological behaviour as the spin models
themselves, by the evidence of the preceding section, seem remarkably well
behaved. We measure
the acceptance rate for the
Metropolis flip move
to confirm that our graphs are really dynamical. The flip can be forbidden
either from the
graph constraints coming from the detailed balance condition or from
the energy change of the spin model, so we can decompose the flip acceptance
rate into two parts:
AL -- the fraction of randomly selected links which can be flipped
satisfying the graph constraints; and
AF -- the fraction of links satisfying the graph constraints which are
actually flipped, i.e. pass the Metropolis test using the Potts model energy
change.
AF and AL are shown for the 2000-node simulations
versus the reduced temperature $t$ in Figs. 2 and 3 respectively.
These are interesting because they show that the dynamical
properties of the phi-cubed graphs are apparently determined by the central
charge
of the models that live on them, even for values of the reduced temperature $t$
some distance from the critical point. The various spin models group into
three separate curves determined by the value of their central charges,
$c=2,4,8$.
The characteristic dip in the flip acceptances away from the phase transition
that is present in the single Potts models and dynamically triangulated
random surface models near the crumpling transition is clearly visible.
For a given $q$ the position of the dips stay roughly fixed as the number of
models is
increased, though they do appear to edge slightly closer to $t=0$ as $q$ is
increased.

We can also examine the distribution of ring lengths in the graph,
which is the discrete equivalent of measuring the distribution of local
Gaussian curvatures
in the continuum.
For pure quantum gravity (no spin model living on the graph)
it is possible to analytically calculate this \cite{15}. The probability $P$
of finding a ring of length $l$ is given by
\begin{equation}
P_{N \rightarrow \infty}(l) = 16 ( {3 \over 16} )^l
{(l-2)(2l-2)! \over l!(l-1)!}
\label{ePG}
\end{equation}
which decays exponentially as $l$ increases. The minimum possible ring length
is 3.
If we plot the fraction of rings of length three (PR3) in Fig. 4
and compare with AF and AL (Figs. 2 and 3) we see that it has a peak very close
to
the dip in AL. This is reasonable since both PR3 and AL
depend only on the graph, whereas AF depends on the Potts model.
The suggestion made in \cite{1}, namely that the peak in PR3 approached
$\beta_c$
from below as $c \rightarrow 1$, is not therefore bourne out by our further
simulations
of multiple models. Just as for the flip acceptances,
the $\beta$ values
of the peaks stay roughly fixed {\it for a given q} as the number of models is
increased, though they get slightly closer to $t=0$ with increasing $q$. It is
also
obvious from Fig. 4 that the PR3 curves
split into three again for $c=2,4,8$
for all the plotted values of $t$, in spite of the small shift in the peak
positions
with $q$.

It is also interesting to examine the {\it height} of the peaks in PR3 (i.e.
maximum values of PR3)
and the values of PR3 at the critical points
as a function of $c$.
If we plot the difference in the fraction of rings of length three at
the critical point $\beta_c$ from the pure gravity
fraction (eq. \ref{ePG} with $l=3$) against the central charge
we find a straight line with slope $0.004(1)$ and intercept $0.004(1)$.
We can also plot a difference using the peak height in the fraction of rings of
length three
against the central charge to obtain another straight line with slope
$0.014(1)$
which passes through the origin.
This peak does not occur at $\beta_c$ so the correlation length is
not infinite and we cannot expect the results of conformal field theory to
apply there, but a linear relation with $c$ persists.
The fits are shown with the data in Fig. 5. This should be compared with a plot
of similar data
for smaller $c$ values in Fig. 7 of \cite{1}, which agrees with the data here
in that range.

We complete our discussion of the graph properties by measuring
their internal fractal dimension $d$.
We use the most naive definition of distance (the fewest links between two
nodes)
so we are considering the ``mathematical geometry'' rather than the ``physical
geometry'' in the terminology of \cite{16} \footnote{It now appears
that the two are, in fact, identical \cite{16a}.}.
We measured the internal fractal dimension at the critical values of $\beta$
for all of
our models on the $N=500,1000,2000$ node graphs and then extrapolated the
results
to graphs of infinite size obtaining the values
listed in the last column of Table 2. From these we can see that the
internal fractal dimension varies little, if at all, with $c$ and $q$ and all
are fairly close to the value of approximately 2.8 that we found in \cite{1}
for single Potts models.
It thus appears that although the curvature distribution may be becoming more
singular (slowly) with increasing $c$, as is witnessed by the increase in PR3,
the fractal dimension is changing very little.

\section{Conclusions}

We have simulated multiple copies of $q=2,3,4$ state Potts models
on dynamical phi-cubed graphs all of which have, at least
naively, $c \ge 1$. We have no evidence for a sudden change in behaviour in the
spin models
as $c$ is increased through 1, but instead found only a gradual change in the
critical exponents we have measured.
The idea that a first order transition could explain the breakdown of the
KPZ/DDK results for
multiple Potts models with $c>1$, as it does for $q > 4$ Potts
models does not appear to be supported by our simulations in which the multiple
Potts copies
behave much as their single brethren do. That the spin models {\it do} interact
with
each other is clear from the change in the exponents and critical temperatures
from those
of the single models. The properties of the phi-cubed graphs such as the flip
acceptances and
the fraction of rings of length three
which vary with the number of models also bear this out, and indicate that the
central charge alone determines the
internal geometry, even for $c>1$.

It is possible that the failure of the KPZ/DDK results for $c>1$ is merely a
breakdown of the ansatz rather
than an indication of a completely new phase. Suggestions to this effect have
been made in \cite{17}
where an extended ansatz involving the coupling of the diffeomorphism ghosts to
the Liouville
theory was suggested. It would be most interesting to explore some of the
suggestions
made for pinning down the parameters of the ansatz in \cite{17} to see if it is
possible
to calculate any of the exponents measured here and to reproduce their slight
variation with $c$.
On the numerical side it would obviously be interesting to extend the work in
the current paper to larger meshes
and better statistics, in particular to attempt to obtain values for $\alpha$
or $\alpha / \nu d$ directly, rather than using the scaling relation. It would
also be a worthwhile exercise to measure the string susceptibility
$\gamma_{str}$ for our multiple
Potts models, as this is one of the easiest exponents to calculate
analytically.

\bigskip
\bigskip
\centerline{\bf Acknowledgements}
\bigskip

This work was supported in part by NATO collaborative research grant CRG910091.
CFB is supported by DOE under contract DE-AC02-86ER40253 and by AFOSR Grant
AFOSR-89-0422.
We would like to thank M.E. Agishtein and A.A. Migdal
for providing us with initial graphs generated by
their two-dimensional quantum gravity code.

\vfill
\eject

\bigskip \bigskip \bigskip
\centerline{\bf Figure Captions}
\begin{description}
\item{Fig. 1.}
Fits to maximum slope of derivative of Binder's cumulant versus $N$
to extract $\nu d$ for a) q=2, b) q=3 and c) q=4.
\item{Fig. 2.}
Flip acceptance AF as function of reduced temperature $t$ for all $q$.
\item{Fig. 3.}
Flip acceptance AL as function of reduced temperature $t$ for all $q$.
\item{Fig. 4.}
Probabilities of rings of length three PR3 as function of reduced temperature
$t$ for all $q$.
\item{Fig. 5.}
Difference in PR3 at $\beta_c$, and at its peak,
from the pure quantum gravity value versus the central charge $c$,
for multiple Potts models.
\end{description}
\end{document}